# Comment on "Laser refrigeration of hydrothermal nanocrystals in physiological media"

Yang Ji†

SKLSM, Institute of Semiconductors, Chinese Academy of Sciences
Beijing 100083, China
†jiyang@semi.ac.cn

The recent report on laser cooling of liquid may contradict the law of energy conservation.

Laser cooling of liquid has recently been reported by Roder *et. al.*[1], however, such a cooling process is unlikely to happen since it violets the law of energy conservation.

The main result of the PNAS paper could be summarized as the following. A laser beam cools a YLF nanocrystal (which is immersed in a liquid) 20 degrees below the room temperature, which then cools the liquid surrounding it. The wavelength of the laser is ~1020nm, the irradiance on the sample is ~25MW/cm$^2$, the diameter of the YLF nanocrystal (10%Yb) is ~1μm, and the size of the cooled region of liquid is ~5μm.

This result may contradict the law of energy conservation, as the cooling power is much less than the thermal conduction power if there is really a temperature gradient around the nanocrystal.

Since the absorption length of the YLF crystal is about a few centi-meters[2], less than 0.01% of the incoming light power could be absorbed by the nanocrystal (whose size is about 1μm). The incoming light power on the sample is about 25MW/ cm² × (1μm)² = 250mW. The cooling power is at most 250mW × 0.01% × 5%=1.3μW, where 250mW is the incoming light power, 0.01% is the efficiency of the absorption, 5%=(1-1000/1020) is the theoretical efficiency of the cooling process (the anti-stokes process). Recap it, the cooling power is at most 1.3μW[3]. Here we have not yet taken into account the rare probability of the anti-stokes process, which could be much less than 1%.

However, the thermal conduction power (heating power) of the surrounding liquid (say, water) is much more than that. Consider a water drop of 5μm radius around the nanocystal. The heating power would be 6mW/cm•deg × 20deg/5μm × 4π × (5μm)²=750μW, where 6mW/cm•deg is the thermal conductivity of water, 20deg/5μm is the temperature gradient, and 4π × (5μm)² is the area of the water sphere.

The estimation above is rough but it should not be far away from the reality. Since the heating power (750μW) is much more than the cooling power(1.3μW), it is unlikely for any cooling process to happen at all. In fact, with the thermal conduction of the surrounding water in mind, it is more likely for the nanocrystal to keep its temperature around the equilibrium (say, ±1 degree), neither above nor below.

It may be worth noting that most of the earlier experiments in laser cooling of solids[4] should be checked for similar issues.

Y. J. thanks Prof. Pauzauskie for clarifying some issues in his paper. Y. J. acknowledges partial support from the NSFC (Grant No. 91321310) and the NBRPC (Grant No. 2013CB922304).